# Integrating AADL and FMI to Extend Virtual Integration Capability

Jérôme Hugues[1], Jean-Marie Gauthier[2], Raphaël Faudou[2]

([1] ISAE-Supaero, [2] Samares-Engineering)

**Abstract** Virtual Integration Capability is paramount to perform early validation of Cyber Physical Systems. The objective is to guide the systems engineer so as to ensure that the system under design meets multiple criteria through high-fidelity simulation. In this paper, we present an integration scheme that leverages the FMI (Functional Mock-Up interface) standard and the AADL architecture description language. Their combination allows for validation of systems combining embedded platform captured by the AADL, and FMI components that represent physical elements, either mechanical parts, or the environment. We present one approach, and demonstrator case studies.

## 1. Introduction

Virtual integration capability is paramount to perform early validation of Cyber Physical Systems. The objective is to guide the systems engineer so as to ensure that the system under design meets multiple criteria through high-fidelity simulation. The general approach is to leverage models as primary artefacts to capture all facets of a system, and tool supports to analyze and simulate the system.

Early study [1] demonstrated the importance of Architecture Description Language to capture many facets of a system, for the embedded perspective, supporting multiple kind of analysis such as timing performance, or safety analysis. Other standards like FMI [2] did the same for the mechanical and control sphere, supporting co-simulation of high-level mechanical models (e.g. built on Modelica) or control command models built around Simulink or SCADE. FMI has been defined with primary objective to support system-level simulations, with a strong emphasis on early validation of systems made of separated models.

A remaining challenge is to combine both spheres. As a matter of fact, architecture description languages capture the organizational structure of the system to be designed, and the exchange of information. The physics is abstracted through devices (sensors and captors) that interact with the environment and mechanical parts. These models capture their behavior as a set of model elements that ultimately lead to equations. FMI allows one to build reusable components from these models, but does not address the construction of simulation itself. This rupture in abstraction makes it difficult to build integrated simulation, and motivates our contribution.

Considering an architectural description model of an embedded system as a primary artifact, we want to streamline the construction of a virtual integration test bench that would integrate external models as environment stimulus, hence allowing through tests of the embedded (or cyber) part. Our contribution builds on the AADL [3] architecture description language, and the FMI standard. Our main contribution in this paper concerns a model-based approach to integrate FMI blocks and AADL so as to leverage existing code generation strategy and build either model-in-the-loop or hardware-in-the-loop simulations.

The paper is organized as follows: in section 2, we review the main technological elements used: AADL and FMI. In section 3, we report on existing FMI-based integration workflows. In section 4 we propose an AADL-based workflow that leads to the generation of simulations. Section 5 is a case study that illustrates the approach. Section 6 proposes future work directions.

## 2. Standards in use

### a. AADL, Architecture Analysis and Design Language

The "Architecture Analysis and Design Language" (AADL) [3] is both a textual and graphical language for model-based engineering of embedded real-time systems. AADL is used to design and analyze software and hardware architectures of embedded real-time systems.

The AADL purpose is to model hardware components (memory, bus, processor, device, virtual processor, virtual bus) and their associated embedded software (data, thread, thread group, subprogram, process). It focuses on the definition of clear block interfaces, and separates the implementations from these interfaces. From the separate description of these blocks, one can build an assembly of blocks that represent the full system. The AADL defines the notion of properties. They model non-functional properties that can be attached to model elements (components, connections, features, instances, etc.). Properties are typed attributes that specify constraints or characteristics that apply to the elements of the architecture such as clock frequency of a processor, execution time of a thread, bandwidth of a bus. As defined, AADL is an Architecture Description Language. Without loss of generality, similar notation such as EAST-ADL or UML/MARTE would provide the same power of expression.

AADL has a rich ecosystem to model, analyze and generate code from models, such as Ocarina [4]. The later aspect is interesting to ease the transition of AADL models' tasks, and communication semantics to an implementation on top of a regular Real-Time Operating Systems. We detail this part in the next section.

Let us note that in such models, the time interval is given by the CPU clock rate (or simulated by a scheduler): AADL models are discrete by nature and fit in the cyber part of cyber-physical systems.

### b. FMI, the Functional Mock-Up Interface

FMI, the Functional Mock-Up Interface, is a standard for the simulation of systems, combining heterogeneous models. The initial revision of FMI mostly focused on models relevant for multi-physical aspects of automotive systems. Since then, it has been widely adopted in several settings, especially for the modelling and simulation of Cyber-Physical Systems, e.g. as part of the Ptolemy project at UC Berkeley [5]. FMI is already supported by an increasing number of tools used in several domains, e.g. Modelica[1] tools, Simulink[2] or SCADE Suite[3]. Through this standard, system designers may mix and co-simulate heterogeneous models built by experts to better understand how a system may be integrated.

FMI, defines an interface to be implemented as a component called FMU (Functional Mock-up Unit). The FMI functions are used (called) by a simulation environment to create one or more instances of the FMU and to simulate them, typically together with other models. An FMU may either embeds its own solver (FMI for Co-Simulation) or requires the simulation environment to perform numerical integration (FMI for Model Exchange). For both approaches, each model is exported in a zip file, called FMU, which contains a binary file of the model and an XML file (named *modeldescription.xml*) which describes the model contents, properties, and interfaces (its associated model variables).

In this paper, we only focus on FMI 2.0 for Co-Simulation, as we are interested in combining self-contained executable blocks and their integration in a system level simulation.

### c. FMI for Co-Simulation

The FMI Standard for Co-simulation is intended to provide an interface standard for coupling two or more simulation tools in a co-simulation environment. Co-simulation is a technique used for the simulation of coupled models.

---

[1] https://www.modelica.org/ [last visited 30/05/2017]
[2] https://fr.mathworks.com/products/simulink.html [last visited 31/05/2017]
[3] http://www.esterel-technologies.com/products/scade-suite/ [last visited 31/05/2017]

A coupled model, is a model that describes a system as a network of (logically or physically) coupled (or connected) components [6, 7]. In the coupled model formalism, the connections between subsystems are represented with connectors, or mathematical equalities. Formally, a coupled model may be represented as a graph structure. For non-causal and continuous models, the graph is undirected. For causal models, the graph is directed. A coupled model is valid if connected ports are compatible regarding their type and causalities [8].

The Figure 1 depicts an FMU represented by a block, with internal state variables x(t), connected to other subsystems of the coupled problem by inputs u(t) and outputs y(t) [9].

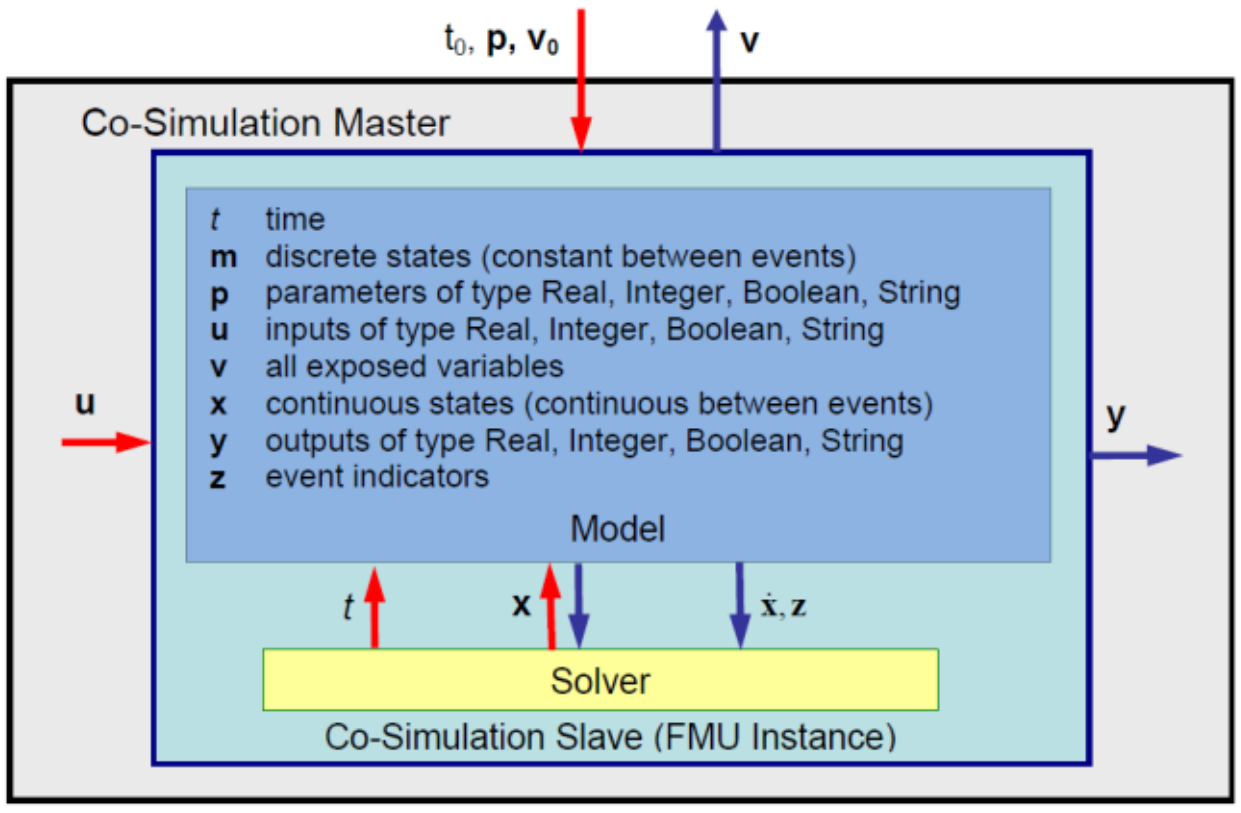

*Figure 1 : Data flow between the environment and an FMU for CS [2]*

As depicted by the Figure 2 there are two possibilities for providing slave subsystems for co-simulation:

- subsystems with their specific solver, they can be simulated in stand-alone mode
- subsystems with the simulation tools in which they have been developed

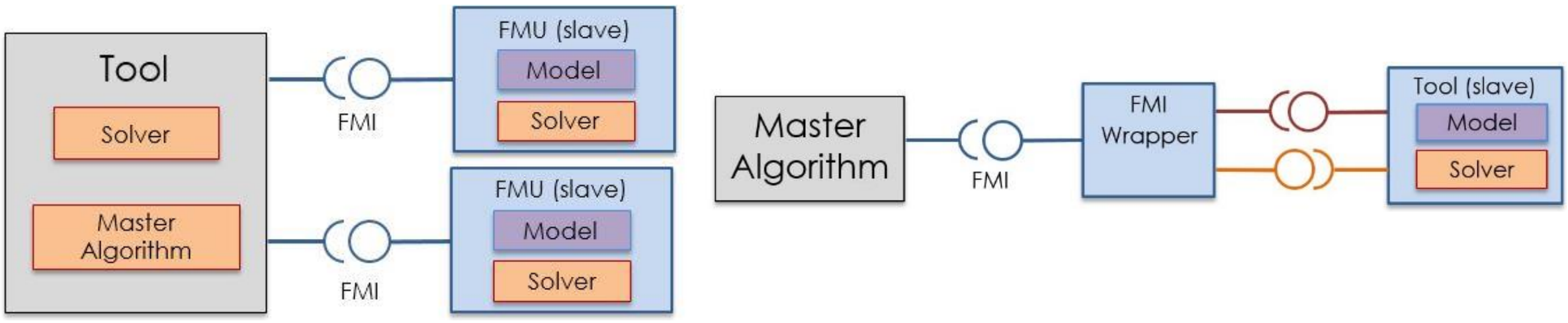

*Figure 2: FMU (CS) integration in stand-alone, and with tool coupling [2]*

The FMI 2.0 specification defines the life cycle (different noticeable states and guarded transitions) of an FMU as the Figure 4. A master algorithm serves several purposes: to instantiate, to initialize, to execute and to synchronize FMUs [10]. An instantiated FMUs is called a slave. Master algorithms synchronize FMUs by controlling the data that are exchanged between FMUs at specific synchronization points called communication steps. The communication step sizes are defined as $hc_i = tc_i+1 - tc_i$, where $tc_i$ are communication points. In the interval between two communication points, the subsystems are solved independently by their respective solvers. Master algorithms control exchanges of data between the subsystems and the synchronization between slaves.

For co-simulation two basic groups of functions should be realized:

1. functions for the data exchange between subsystems,
2. functions for algorithmic issues to synchronize the simulation of all subsystems and to proceed in communication steps $tc_i \rightarrow tc_{i+1}$ from initial time $t_{c0} = t_{start}$ to end time $t_{cN} = t_{stop}$.

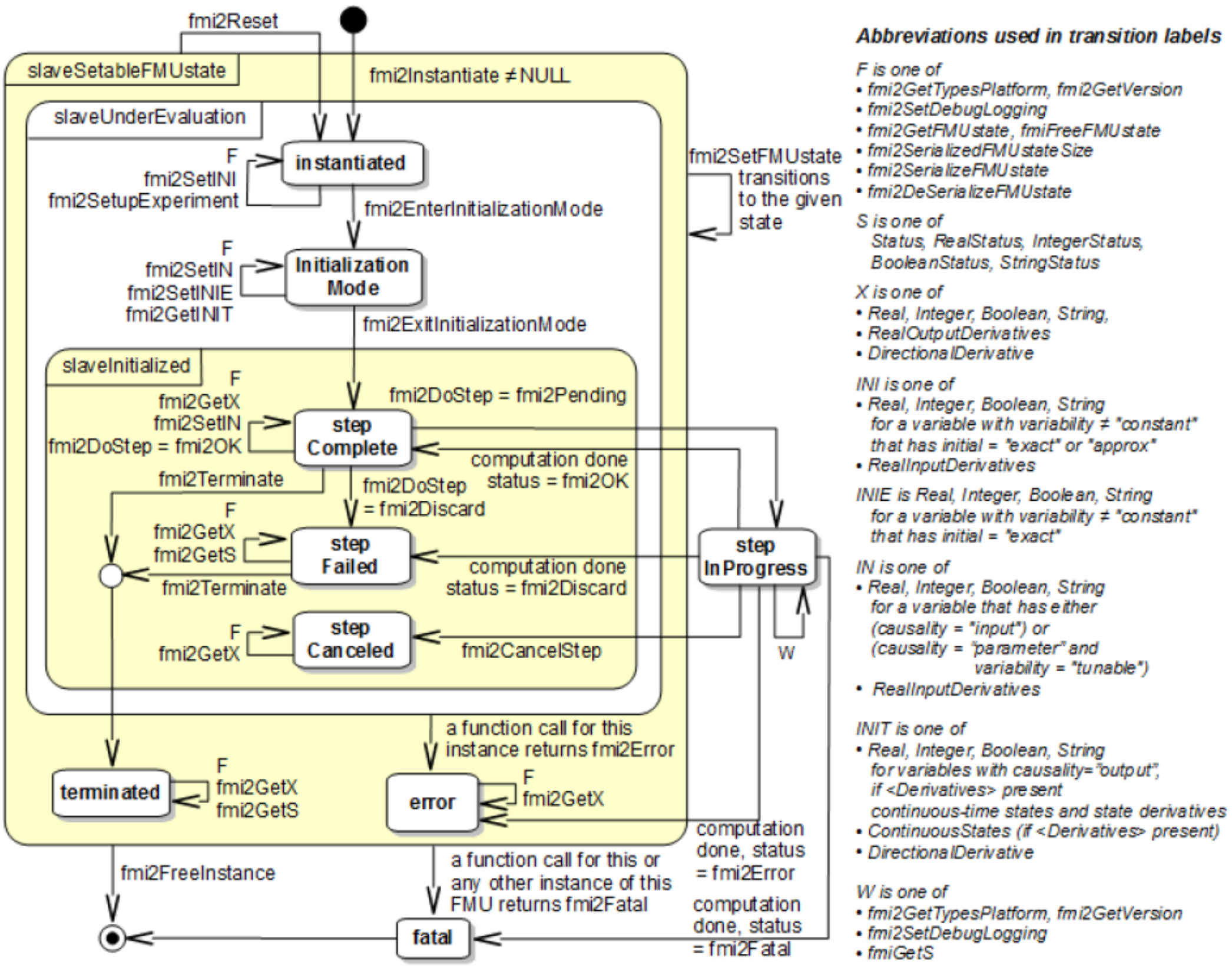

*Figure 3: State-machine of FMI Co-Simulation [2]*

### d. Combining ADL and FMI – benefits and challenges

An Architecture Description Language (ADL) model captures the organizational structure of the system, along with executable blocks that represents its inner structure. It is fully configured so as to faithfully represent its behavior in terms of timing and communication. Several studies explored the capability to perform model checking or analysis of such model. Yet, a usual limit is the difficulty to capture a model of the environment that would act as a stimulus for the verification part.

FMI has been designed to support system simulation. We claim it can also be used also for verification of real-time embedded systems, by combining a mock of the environment as an FMU and the system under test. In this paper, we illustrate our approach to address this part using AADL.

Combining AADL and FMI would equip system architects with a tool to prepare in advance integration phases through a virtual test bench. Indeed, this would make easier the early validation of AADL models: the stimulus provided by FMUs (that represent the physical environment) could be considered as test cases for AADL models. Using FMU would also give access to models with high-level of fidelity, with connections to other engineering models beyond a naïve model of a plant.

However, this combination of discrete-time execution semantics for the cyber part (AADL) with systems dynamics (FMI, multi-physics simulation) is not an easy task. Such hybrid models raise issues regarding time management in co-simulation, typical issues concern the time step used to synchronize elements, or intermediate extrapolations performed by each model when no inputs are exchanged. In [11], the author provides a thorough review of these topics in the general case.

In the following, we review existing work prior to propose an integrated workflow in section 4.

## 3. Related Work

FMI was first designed to co-simulate physical and hybrid systems (continuous and discrete) specified using DAEs (Differential and Algebraic Equations) and discrete events. In this paper, we focus on the use of FMI in the context of critical embedded software design with AADL. We plan also on leveraging existing code generation from AADL that targets Real-Time Operating Systems (RTOS).

In the following, we present related work that a) integrate FMI with modelling language, b) that combine FMI with RTOS code, and c) that use FMI for hybrid systems co-simulation.

### a. Integrating FMI with high-level modelling language

The integration of FMI with modelling language such as UML or SysML is an addressed challenge. In [12], the authors proposed a co-simulation environment that combines the execution of UML models (with fUML[4]) and FMI within the Moka Papyrus plugin. This work the co-simulation of hybrid systems, e.g. a controller modelled as an activity diagram (discrete), and its environment as physical and continuous FMUs. Feldman et al. [13] proposed to export Rhapsody SysML blocks into FMUs, with a limitation on flow ports and attributes (the behavior of SysML block, e.g. state-machines or activities, is not supported for now).

Combining EAST-ADL and FMI has also been investigated during the MAENAD project [14]. The work resulted in FMI 1.0 import capability within EAST-ADL model using model transformation technology. However, this work focused more on the semantic mapping between EAST-ADL and FMI, than on co-simulation issues and execution. Let us note current FMI technology lacks means to build simulation assemblies. The standard System Structure and Parameterization (SSP) [15] will complement FMI on this particular topic.

As our work concentrates at a lower level of CPS design, i.e. embedded software modelling and simulation using AADL, we propose to complete these related works at the lower left-side of the V-cycle with RTOS validation capabilities using FMI.

### b. Combining FMI with RTOS code

In [16], the authors propose to adapt embedded software to comply with FMI for co-simulation. More precisely, the authors propose to advance the clock of the RTOS, by overwriting the idle thread and waiting for a signal to start execution. Pohlmann et al. [17], proposed to generate FMUs from UML software specification, where the clock is specified in a DSL named MechatronicUML. This clock is used to measure execution time and to specify Real-Time properties within timed state-machines.

On the contrary, we aim at validating a system from its ADL model. Hence, we propose to simulate the behavior of the embedded processor, which executes the target code. Hence, one can perform co-simulation of CPS, closest to the actual implementation without the need of specific hardware (between Software-In the Loop and Hardware in the Loop), or on the final target. This is left as a late-binding decision.

### c. FMI for hybrid co-simulation

Co-simulating discrete (software) and continuous models (physical), raises several issues encountered during the study and experiments. Indeed, mixing continuous and discrete behavior in a co-simulation framework is not well handled by FMI. The representation of time and its management are the key issues of FMI based co-simulation approaches. In Cremona et al. [18], the authors identified extensions to FMI for supporting hybrid co-simulation: use integer time instead of floating point time representation, automatic choice of time resolution, use of "super-dense" time, and the use of absent signal. The proposed solution of [18] satisfies all the requirements for hybrid co-simulation stated in [19]. Unfortunately, it imposes strong constraints on the usability: these extensions are not backward-compatible with existing FMI 2.0 FMUs.

[4] http://www.omg.org/spec/FUML/1.2.1/ [last visited 02/06/2017]

Finally, using the dependency graph as an asset to get more precise hybrid co-simulation results is an issue investigated in several works, especially in [20], whose authors propose to generate master algorithm based on the dependency graph (with and without loop) and on the step-size of each FMU (multi-clock management). We are also interested in the recent results of DACCOSIM [21], which propose to generate master algorithms for parallel and distributed co-simulation using hierarchical FMUs.

## 4. Integrating FMUs as AADL blocks

In this Section, we present the integration workflow of FMU components within AADL models. We view the integration of FMU as an integration process. Starting from an AADL model, one aims at integrating FMU as an executable block, similar to the inclusion of other functional models in AADL, such as C, Ada, Scade, Simulink that are already supported by our AADL toolchain Ocarina. Each block is integrated as a subprogram block, that is triggered by its enclosing component such as a thread or a device.

Ocarina[5] [19] is a model processor for the AADL. It supports code generation targeting a wide variety of RTOS (RTEMS, RT-POSIX, FreeRTOS, ARINC653). It maps AADL constructs onto the PolyORB-HI runtime that abstract RTOS constructs. It preserves the initial semantics of the AADL model.

### a. Integration of FMU in AADL workflow

First, let us say that one receives an FMU that models and simulates the mechanical part (physical) of a larger system. This FMU should be integrated with a controller (cyber part), designed with AADL. The goal is to verify that the controller behaves as expected. The integration workflow is presented Figure 4. The first step consists in an automated translation of the FMU as AADL model using an algorithm, which 1) unzips the FMU file, 2) parses the *modeldescription.xml* file to create AADL elements respecting the mapping of Table 1, and 3) creates the FMI wrapper as set of AADL constructs: subprogram that capture the execution (entrypoint of the simulator) and corresponding C implementation, thread and device abstractions. Then, one could connect the FMU with the larger AADL model to build a coupled model.

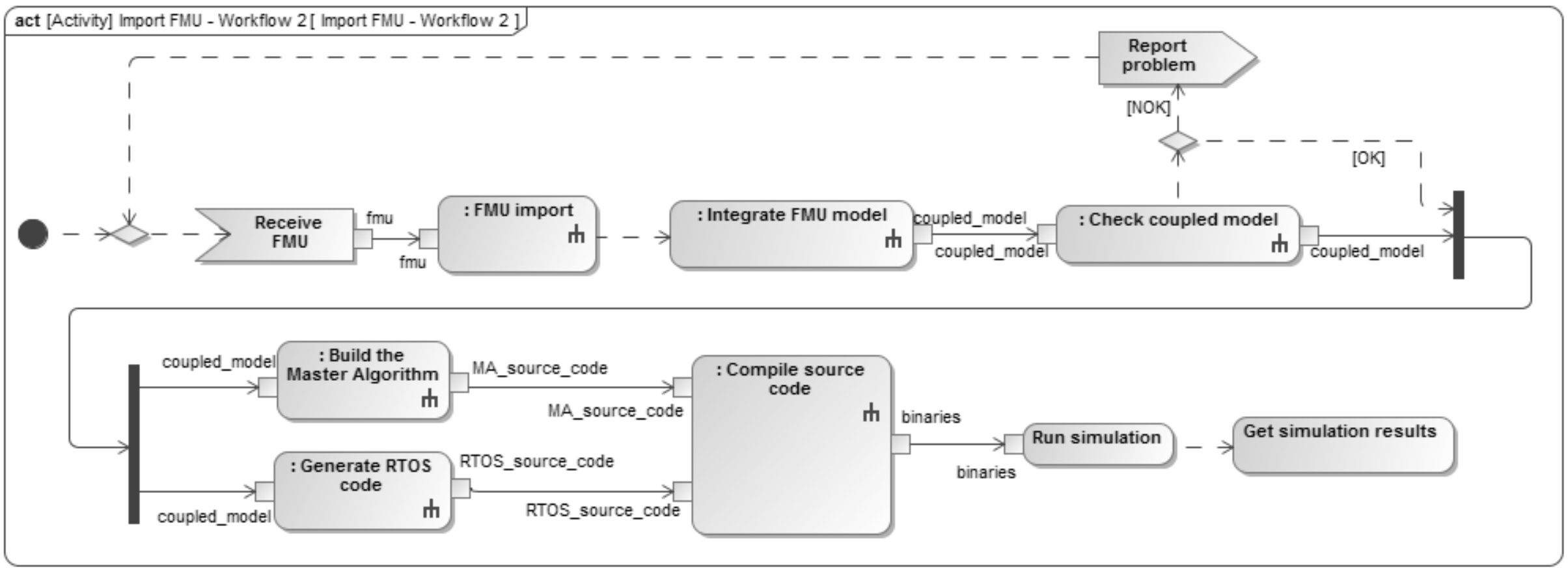

*Figure 4: Integrating an FMU as an AADL component – Workflow*

### b. Integration of FMU execution semantics

FMU-based simulations rely on the concept of a Master Algorithm that orchestrates the overall simulation at each time step. Then, the FMUs perform a calculation step, and the resulting values are propagated to the scheduler, which launches tasks depending on the overall simulation time. The Master Algorithm can be built based on the dependency graph of the coupled model.

5 http://openaadl.org

However, in the case of algebraic loop detection, the calculation of the dependency graph is aborted, and a generic Master Algorithm is provided. In parallel, the target RTOS code is generated from the AADL model, using Ocarina. Finally, the whole code is compiled and linked to be executed.

*Table 1: Mapping Between FMI and AADL*

| Concepts | FMI | AADL |
|---|---|---|
| Component | FMU | Subprogram/Device |
| Input/Output port | Causality In/Out | In port / Out port |
| Discrete port / Continuous port | Variability Discrete/Continuous | Event port / Data port |
| Types | Real, Integer, Boolean | Base_Types::Float,<br>Base_Types::Integer,<br>Base_Types::Boolean |

As stated by Table 1, FMU blocks are mapped to AADL device, that is an abstraction of a device interacting with the physical world captured by the FMU model. Hence, FMU interactions are explicitly discretized by the activation of the device by other part of the model, e.g. reading from a sensor will trigger the corresponding FMU at the corresponding sampling time. Hence, one can integrate either continuous time or discrete time in the simulation.

In our approach, FMUs are embedded inside an AADL model that already have an execution semantics, and a scheduler. As a result, the Master Algorithm is implicitly defined by the combination of the scheduling parameters of all blocks: signals are captured in AADL event port communications; data propagation as AADL data port communications; scheduling is controlled by the scheduler of the system, e.g. priority-driven scheduler. Thus, the cyber part of the system interacts with the environment through polling (periodic read), interrupts (reception of events from the environment) or actuation. These are captured by corresponding port directions in the AADL model.

As a consequence, the co-simulation time is linked to scheduler time. We used signals to start the scheduler until the communication time step is reached. The implementation of the master algorithm is automatically generated using the Ocarina AADL code generator by translating AADL tasks and communication ports to the corresponding C artefacts. Thanks to the versatility of the code generation process, we can either really on true (wall clock) time, or simulated time using a simulator of a RTOS. Hence, this approach allows seamless integration of FMU as functional models.

## 5. Case Study and Experiment Results

In this section, we list two case studies built on the previous integration workflow[6].

### a. Moonlander

A Moon Lander model [22] was used to investigate co-simulation's time and scheduler's time issues. We built a Modelica model of the physical model and exported it into an FMU 2.0 for Co-Simulation. The controller implements a basic strategy to control the descent of the vehicle, and triggers the thrusters. A first version has been implemented in Modelica to check the correctness of the controller, and simulated in the OpenModelica framework.

Then, the FMU of the physical model was imported as an AADL component following the semantics mapping of the Table 1, and connected to a reimplementation of the controller in C. This controller and the plant, seen as an AADL device that samples the environment, have been connected in an AADL model (see Figure 6). This model indicates the data types exchanges and the scheduling parameters of the controller.

[6] Other case studies are available through http://www.openaadl.org

From the system designer perspective, the AADL model captures the configuration of the controller, and a device that interacts with the environment as a regular device driver. Here, FMI plays its role of "mock-up", instead of connecting this model to the implementation of a driver, we connect it to the model that simulated the environment it interacts with.

To simulate the overall system, we generated the C code of the AADL controller with Ocarina, along with code archetype for the various tasks and communication channels. We linked it to the FMU as an external library. We rely on the FMUSDK2 from Modelon (adapted by University of California – Berkeley[7]) to build a generic entry point to load FMUs and to compute simulation steps triggered by the host process. For this experiment, we relied on the GNU/Linux OS, combined with RT-POSIX API call to implement a real-time behavior for the controller.

Through analysis of the execution logs, we could assess the simulation has the same behavior for both the FMI-coupled model and the Modelica model.

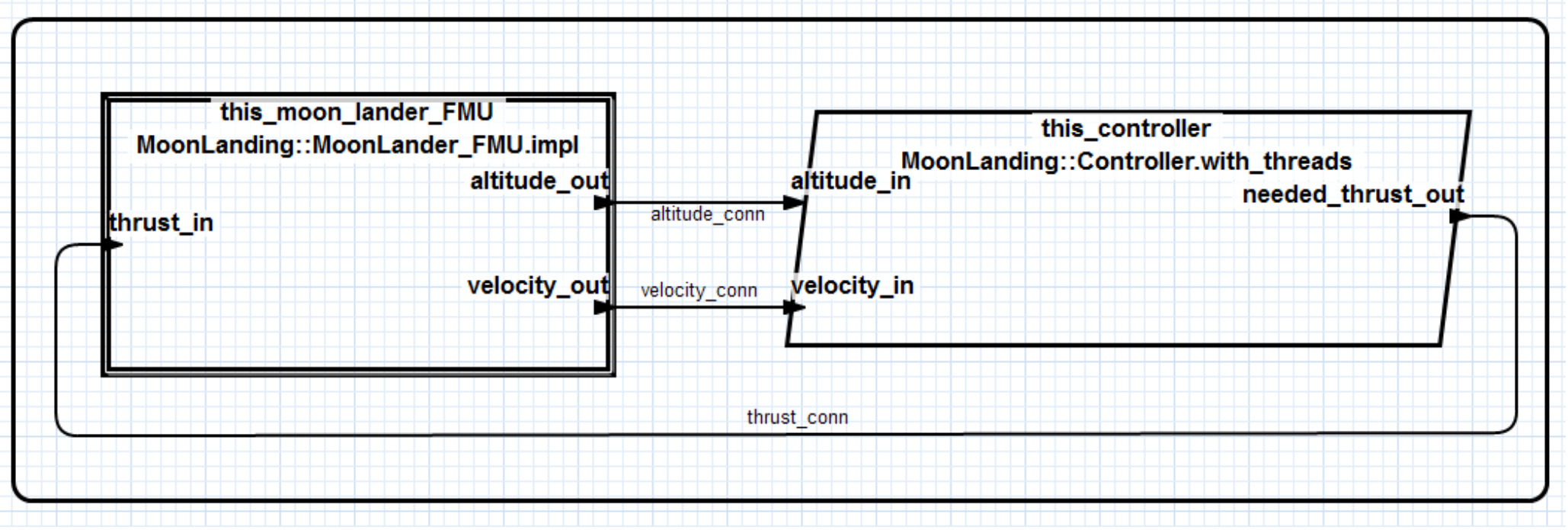

*Figure 5 AADL model of the lunar lander*

### b. ROSACE case study

The ROSACE case study [23] was used to investigate co-simulation graph issues, but also scalability. ROSACE has multiple implementations in C/POSIX, C/ARINC653, Simulink Giotto or Ptolemy. It is a reference benchmark for CPS simulation.

The environment of the controller is composed of three FMUs: an engine, an elevator, and the aircraft dynamics. These FMUs are connected to exchange physical quantities, and to the controller that is modeled in AADL. This controller is made of 11 periodic tasks interconnected. This model has a higher level of complexity compared to the MoonLander case study, with more FMUs and tasks.

To build these FMUs, we modelled the environment using the Modelica language and we generated the FMUs with JModelica. The controller has been implemented in Simulink, and later translated in C. Then, based on [20, 21] we have investigated the construction of the overall dependency graph to generate automatically the Master Algorithm. This algorithm has been translated as a set of AADL scheduling configuration parameters. The models are simple enough to be full discretized, thus a causal graph can be deduced to capture the whole simulation behavior.

We could leverage the ROSACE validation script to ensure that our simulation was also consistent with other simulations done either in Simulink, Ptolemy/HLA or Giotto.

### c. Lessons learnt

Through these two case studies, we could generate virtual integration test bench for cyber physical systems. We demonstrated the capability to connect ADL model with models simulating the environment using the FMI framework. This is a first step towards full generation of simulation environment.

---

[7] https://github.com/cxbrooks/fmusdk2 [last visited 03/06/2017]

The architectural description of the system has been demonstrated to be enough to interact with the FMU-based environment blocks. This is a consequence of using causal systems: one can simulate the environment up to the instant required by the cyber part.

## 6. Conclusion and Further Work

In this paper, we addressed the early validation of embedded systems. We proposed a general approach to bind architectural description, amenable to code generation, to FMU blocks. This enables the construction of virtual integration test bench. First, we presented the various elements of contexts and related work. Then, we illustrated how FMI blocks can be bound to AADL models so to serve as a mock of the environment seen through a device. Hence, one can test an AADL model considering a representative model of the environment. Leveraging high-fidelity model turned into FMU, one can test more precise interaction scenario. Future work activities will increase the number of case studies, to stress timeliness issues, e.g. multi-clock scenarios. Another aspect will consider integrating multiple simulators like Instruction Set Simulators for precise simulation of hardware blocks, and interoperability with domain-specific simulators to simulate the occurrence of faults and defects.